\documentclass{Interspeech2024}

\usepackage{adjustbox}
\usepackage{graphicx}
\usepackage{multirow}
\usepackage{cite}
\usepackage{colortbl}
\usepackage{xcolor}

\interspeechcameraready

\title{CodecFake: Enhancing Anti-Spoofing Models Against Deepfake Audios from Codec-Based Speech Synthesis Systems}

\name[affiliation={1,2}]{Haibin}{Wu}
\name[affiliation={1,2}]{Yuan}{Tseng}
\name[affiliation={1,2}]{Hung-yi}{Lee}

\address{
  $^1$Speech Processing and Machine Learning Laboratory, National Taiwan University \\
  $^2$Graduate Institute of Communication Engineering, National Taiwan University}
\email{f07921092@ntu.edu.tw}


\begin{document}

\maketitle

\begin{abstract}

Current state-of-the-art (SOTA) codec-based audio synthesis systems can mimic anyone's voice with just a 3-second sample from that specific unseen speaker. 
Unfortunately, malicious attackers may exploit these technologies, causing misuse and security issues.
Anti-spoofing models have been developed to detect fake speech. 
However, the open question of whether current SOTA anti-spoofing models can effectively counter deepfake audios from codec-based speech synthesis systems remains unanswered.
In this paper, we curate an extensive collection of contemporary SOTA codec models, employing them to re-create synthesized speech. 
This endeavor leads to the creation of CodecFake, the first codec-based deepfake audio dataset. 
Additionally, we verify that anti-spoofing models trained on commonly used datasets cannot detect synthesized speech from current codec-based speech generation systems.
The proposed CodecFake dataset empowers these models to counter this challenge effectively.

\end{abstract}

\section{Introduction}
\label{intro}

Neural audio codecs were first proposed to allow better compression of audio data, thus reducing file storage sizes and improving transmission latency significantly.
Researchers have also found that neural audio codecs serve as effective tokenizers for converting continuous speech into discrete codes suitable for autoregressive speech generation tasks, akin to the role of language modeling in natural language processing. 
Consequently, the past three years have seen substantial development of codec-based audio synthesis systems \cite{wu2024towards,borsos2023audiolm,rubenstein2023audiopalm,agostinelli2023musiclm,wang2023neural,zhang2023speak,wang2023viola,yang2023uniaudio,chen2023lauragpt,wang2023speechx,copet2023simple,lan2023stack,kreuk2022audiogen}. 
Current state-of-the-art codec-based audio synthesis systems \cite{wu2024codec,wang2023neural,zhang2023speak,wang2023viola,yang2023uniaudio,chen2023lauragpt,wang2023speechx} have advanced to the point where they can accurately mimic someone's voice with just a 3-second recording from an unseen speaker. 
Unfortunately, these advancements raise concerns about potential misuse and security issues, as malicious attackers may exploit these technologies to impersonate others' speech.

The ASVspoof challenge, as cited in \cite{yamagishi2021asvspoof,todisco2019asvspoof,kinnunen2017asvspoof,wu2015asvspoof}, serves the purpose of attracting international teams to participate in mitigating synthetic speech using anti-spoofing models.
Anti-spoofing models are designed to perform binary classification tasks, specifically identifying and discarding fake audio.
In recent years, there has been remarkable progress in the development of high-performance anti-spoofing models \cite{yamagishi2021asvspoof,todisco2019asvspoof,kinnunen2017asvspoof,wu2015asvspoof,das2020attacker,wu2023defender, Yi2022ADD} for traditional speech synthesis systems. 
However, codec-based speech synthesis systems are relatively new, 
and whether current state-of-the-art anti-spoofing models can effectively counter deepfaked audios generated by codec-based speech synthesis systems remains an open question.

In this paper, we assemble an exhaustive collection of state-of-the-art contemporary codec models, employing them to re-synthesize speech.
The collection includes six diverse mainstream codec papers, each characterized by distinct training implementations, yielding a total of 15 unique codec models.
This undertaking results in CodecFake, which marks the inception of the first-ever codec-based deepfake audio dataset.
\footnote{We note that a concurrent preprint with similar ideas \cite{xie2024codecfake} was submitted to arXiv during the Interspeech 2024 anonymity period.}
Furthermore, we conduct a comprehensive analysis, carefully assessing the effectiveness of cutting-edge anti-spoofing models when confronted with synthesized speech generated by cutting-edge neural audio codec models.
In addition, we also leverage the CodecFake dataset to create enhanced anti-spoofing models for detecting codec-based deepfaked audios.

We find that: 
(1) Models trained on the ASVspoof dataset struggle to detect not only the synthesized speech from codec models but also from current codec-based speech generation models.
(2) Our proposed CodecFake dataset effectively enhances models' ability to tackle synthesized speech from current state-of-the-art codec-based speech generation systems (equal error rate 0.4\% for synthesized speech from VALL-E, a state-of-the-art codec-based speech generation model).

We will release all code and data \footnote{https://codecfake.github.io}, and we aspire that this study and the CodecFake dataset will contribute to mitigating malicious attacks stemming from current advanced codec-based speech synthesis systems.

\begin{figure}[tb]
  \centering
  \includegraphics[trim={4.7cm 0 4.7cm 0},clip,width=0.95\linewidth]{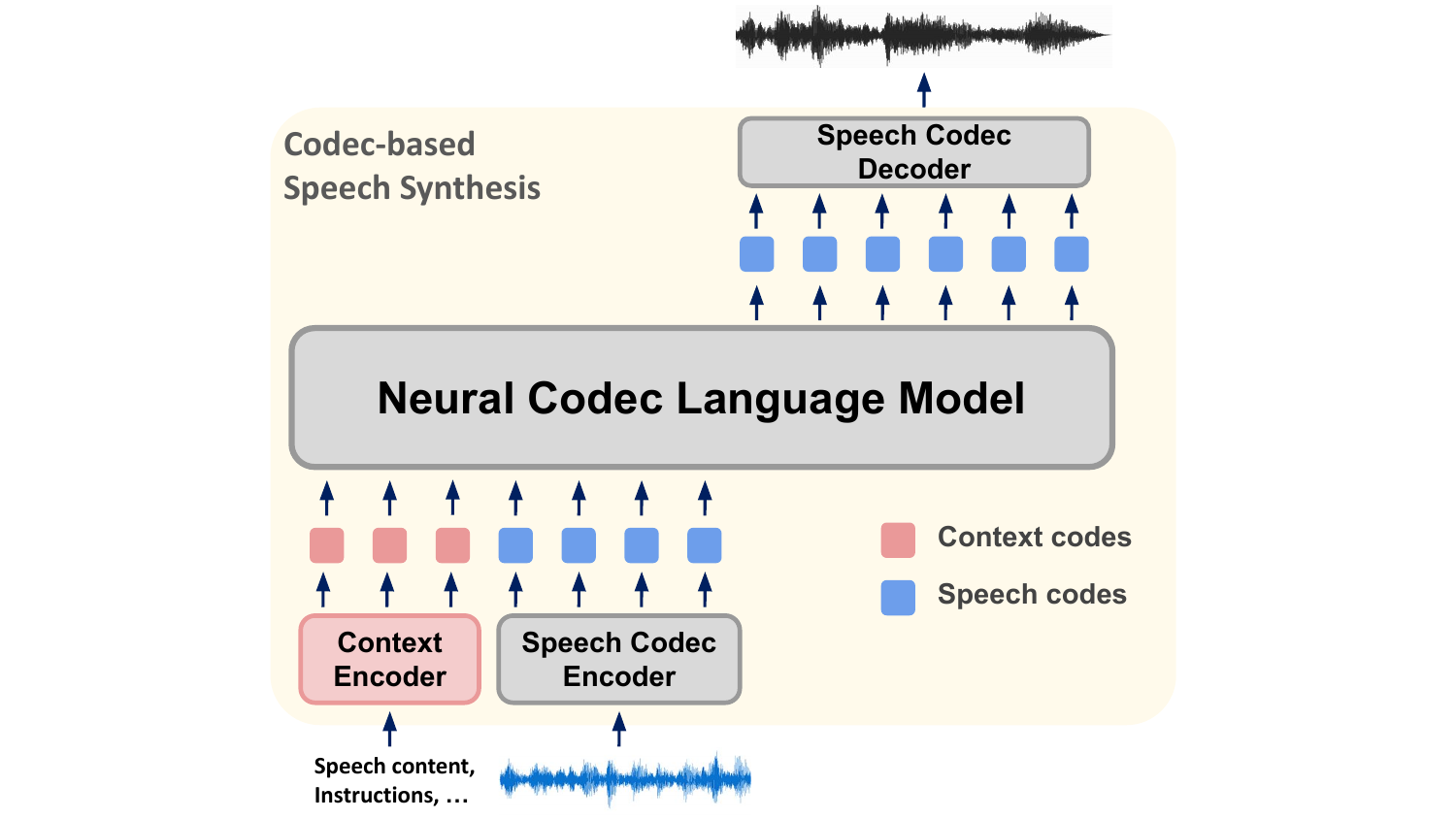}
  \caption{
  Codec-based speech generation pipeline.
  }
  \vspace{-3mm}
  \label{fig: framework}
\end{figure}

\section{CodecFake}
\label{method}

This section presents the dataset creation procedure, design principles, and the rationale behind the CodecFake dataset.

\subsection{Codec-based speech generation}
As depicted in Figure~\ref{fig: framework}, codec-based speech generation starts with converting context information, such as the desired speech content or textual instructions, into discrete context codes while encoding audio into discrete speech codes at the same time.
These context and speech codes are then fed into the codec language model to generate a target speech code sequence.
Lastly, the target speech code sequence is sent to the speech codec decoder to synthesize the final output waveform.

Various designs exist for codec language models (center gray block in Figure~\ref{fig: framework}) \cite{borsos2023audiolm,rubenstein2023audiopalm,agostinelli2023musiclm,wang2023neural,zhang2023speak,wang2023viola,yang2023uniaudio,chen2023lauragpt,wang2023speechx,copet2023simple,lan2023stack,kreuk2022audiogen}, resulting in numerous variations.
To mitigate the impact of these variations and preserve the sole invariant factor of the codec-based speech generation, we exclusively utilize the codec resynthesis procedure to generate our proposed CodecFake dataset.
This process involves encoding speech into discrete codes and then decoding these codes back into speech.
Experimental results in Table~\ref{tab: tts} verify that the CodecFake dataset, generated according to this principle, enhances anti-spoofing models to catch deepfaked speech produced by state-of-the-art codec-based speech synthesis systems.

\subsection{Codec adoption}
To ensure our generated dataset achieves extensive coverage and comprehensiveness, we systematically gather all current open-source codec models spanning six distinct types, each with its unique training settings. 
This effort yields a total of 15 distinct codec models, summarized in Table~\ref{tab: codec_info}.

\begin{table}[t!]
    \renewcommand{\arraystretch}{1.2}
    \setlength\tabcolsep{4pt} 
    \caption{Details of adopted neural audio codec models.
    "A-F" represent different codec frameworks, 
    $N_c$ represents the number of codebooks,
    SR represents audio sampling rate (in kHz),
    and BPS represents bits per second.
    }
    \centering
    \begin{adjustbox}{width=1\columnwidth}
    \begin{tabular}{c ccccc}
    \toprule
    & \multirow{2}{*}{Name} 
    & \multirow{2}{*}{\begin{tabular}{c} Training \\ Data \end{tabular}} 
    & \multirow{2}{*}{$N_c$} 
    & \multirow{2}{*}{SR}%
    & \multirow{2}{*}{BPS} \\
    & & & & &  \\
    \midrule
    A & 16k & Librispeech & 8 & 16 & 4 \\
    \midrule
    B1 & hifi\_16k\_320d & LibriTTS & 4 & 16 & 2 \\
    B2 & hifi\_16k\_320d\_large\_uni & VCTK & 4 & 16 & 2 \\
    B3 & hifi\_24k\_320d & AISHELL & 4 & 24 & 3 \\
    \midrule
    C & 24k\_320d & Valentini  & 8 & 24 & 6.4 \\
    \midrule
    D1 & 16k & Common Voice, DAPS & 12 & 16 & 6 \\
    D2 & 24k & MUSDB, VCTK & 32 & 24 & 24 \\
    D3 & 44k & AudioSet, Jamendo & 9 & 44.1 & 8 \\
    \midrule
    &  & Common Voice &  &  &  \\
    E & 24k & DAPS, Jamendo  & 8 & 24 & 1.5 \\
    &  & AudioSet, FSD50K  &  &  &  \\
    \midrule
    F1 & en\_libritts\_16k\_gr1nq32ds320 &  & 32 & 16 & 16 \\
    F2 & en\_libritts\_16k\_gr8nq32ds320  & LibriTTS & 32 & 16 & 16 \\
    F3 & en\_libritts\_16k\_nq32ds320 & Subset & 32 & 16 & 16 \\
    F4 & en\_libritts\_16k\_nq32ds640 &  & 32 & 16 & 8 \\
    F5 & zh\_en\_16k\_nq32ds320 & In-house, 25k hrs & 32 & 16 & 16 \\
    F6 & zh\_en\_16k\_nq32ds640 & (en \& zh-cn) & 32 & 16 & 8 \\
    \bottomrule
    \end{tabular}
    \end{adjustbox}
    \label{tab: codec_info}
\end{table}

\begin{table}[t!]
    \renewcommand{\arraystretch}{1.2}
    \setlength\tabcolsep{4pt} 
    \caption{
    Statistics of various data sources used.
    }
    \centering
    \begin{adjustbox}{width=1\columnwidth}
    \begin{tabular}{c cccc}
    \toprule
    & \multirow{2}{*}[-3pt]{Data Source} 
    & Training %
    & Validation
    & Testing \\
    \cmidrule(lr){3-5}
    & & \multicolumn{3}{c}{\# utts./ \# spks.} \\
    \midrule
    \multicolumn{2}{l}{\textit{Dataset Generation}} & & & \\
    CodecFake      & VCTK          &  42752 / 103   & 735 / 2  & 755 / 2 \\
    VCTK           & VCTK          &  42752 / 103   & 735 / 2  & 755 / 2 \\
    \midrule
    \multicolumn{2}{l}{\textit{Zero-shot Evaluation}} & & & \\
    ASVspoof   & ASVspoof 2019 & N/A & N/A & 71237 / 67 \\ %
    VALL-E     & VCTK          & N/A & N/A & 755 / 2 \\
    VALL-E X   & Demo Page     & N/A & N/A & 36 / 36 \\
    SpeechX    & Demo Page     & N/A & N/A & 12 / 12 \\
    \bottomrule
    \end{tabular}
    \end{adjustbox}
    \label{tab: testing_data_info}
\end{table}

\begin{figure*}[t!]
  \centering
  \includegraphics[trim={1cm 5cm 0 0},clip,width=1.0\textwidth]{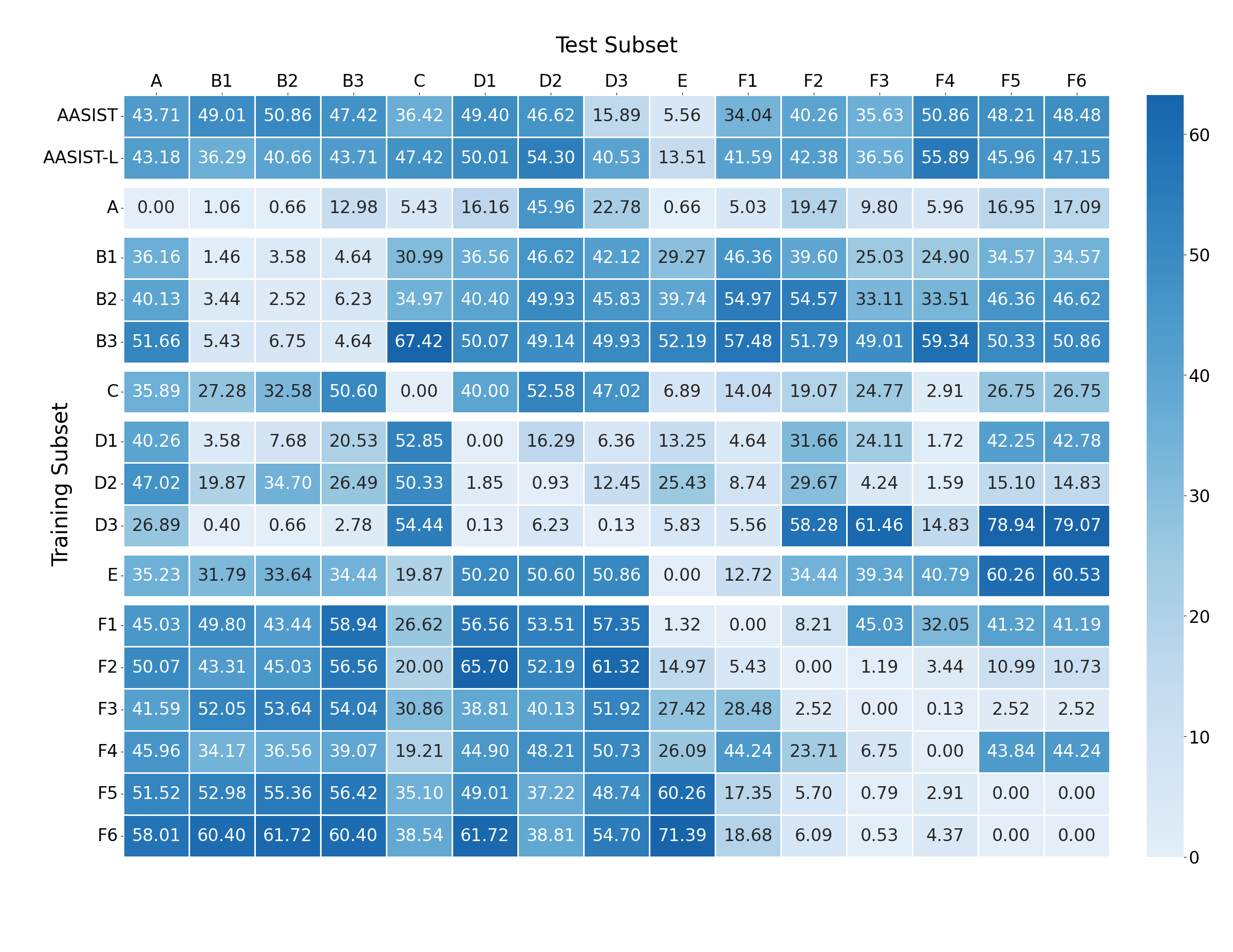}
  \caption{
  Cross-testing results of anti-spoofing models on CodecFake test subsets. 
  We report equal error rate (EER), where lower is better.
  Row AASIST and Row AASIST-L are baselines trained on the ASVspoof 2019 logical access training set using different architectures,
  while rows A-F6 are AASIST-L models trained on one training subset of CodecFake.
  Columns A-F6 indicate the specific CodecFake test subset evaluated.
  }
  \vspace{-5mm}
  \label{fig: codecs_t}
\end{figure*}

\noindent
\textbf{Encodec} (E) \cite{defossez2022high} stands as one of the pioneering implementations of neural codec models, embodying a classic neural codec architecture comprising encoder, quantizer, and decoder modules.
It further augments its capabilities by integrating supplementary LSTM layers and harnessing a Transformer-based language model to model residual vector-quantized codes, thereby amplifying its sequence modeling performance.

\noindent
\textbf{SpeechTokenizer} (A) \cite{zhang2023speechtokenizer} is a unified speech tokenizer designed for speech-language models. 
Similar to Encodec, it also implements an encoder-decoder architecture enhanced with residual vector quantization (RVA).
By integrating semantic and acoustic tokens, SpeechTokenizer hierarchically separates various facets of speech information across different RVQ layers. 
Specifically, SpeechTokenizer is designed to regularize the first RVQ layer to learn Hubert tokens \cite{hsu2021hubert}.
The authors claim that employing such techniques can lead to improved disentanglement of information across various RVQ layers.

\noindent
\textbf{AcademiaCodec} (B) \cite{yang2023hifi} proposes a group-residual vector quantization technique, tailored explicitly for generation tasks. 
It aims to enhance the reconstruction performance using a limited number of codebooks, consequently achieving an impressively low bit rate per second (BPS). 
This low BPS is of utmost significance as it effectively addresses the challenge
of lengthy speech tokens in speech-language modeling, resulting in reduced sequence lengths.

\noindent
\textbf{AudioDec} (C) \cite{wu2023audiodec} represents an enhanced version of Encodec, 
implementing a group convolution mechanism to facilitate real-time operation of the streamable network, 
while also harnessing the capabilities of HiFiGAN \cite{kong2020hifi} to effectively generate high-fidelity audio at a sampling rate of 48 kHz.

\noindent
\textbf{Descript Audio Codec} (D) \cite{kumar2023high} combines a multitude of training techniques, such as periodic activation functions \cite{ziyin2020neural}, 
enhanced residual vector quantization using factorized and L2-normalized codes, 
random quantizer dropout to preserve audio reconstruction quality, 
as well as refining adversarial and reconstruction loss during the training process. 
Out of all the techniques employed, they emphasize the pivotal role played by the periodic activation function.

\noindent
\textbf{FunCodec} (F) \cite{du2023funcodec} proposes a frequency-domain codec. 
The authors claim they can achieve comparable performance with fewer parameters and lower computation complexity. 
Meanwhile, they also find incorporating semantic information in the codec tokens improves speech quality at low bit rates.

\subsection{Dataset generation pipeline}
The most famous anti-spoofing dataset, the ASVspoof 2019 database \cite{todisco2019asvspoof}, is based on the Voice Cloning Toolkit (VCTK) corpus \cite{vctk}.
Following the methodology of ASVspoof 2019, we generate the CodecFake dataset based on the VCTK corpus. 
The set of 107 speakers is divided into three speaker-disjoint sets for training, validation, and testing. 
Speakers p226 and p229 are allocated for generating validation data, while speakers p227 and p228 are reserved for generating testing data. 
The remaining speakers are used to generate training data.

We use 6 different neural audio codec frameworks, with a total of 15 different pre-trained codec models to re-synthesis speech based on VCTK.
As shown in \ref{fig: framework}, a typical neural audio codec model consists of an encoder and a decoder.
Using a pre-trained codec encoder, we first encode real speech into discrete codes.
With these discrete codes as input, we re-synthesize speech using each respective codec decoder.
CodecFake contains 15 subsets, generated by an existing pre-trained neural audio codec model in Table~\ref{tab: codec_info}.
Each subset shares the same statistics as VCTK, with training, validation, and testing data.
Statistics are shown in \textit{Dataset Generation} part of Table~\ref{tab: testing_data_info}.

\begin{table*}[t!]
    \renewcommand{\arraystretch}{1.3}
    \setlength\tabcolsep{3.5pt} 
    \caption{
    Comparison of fake speech detection performance (EER) of models trained on ASVspoof and CodecFake datasets.
    Models that outperform those trained on ASVspoof (\textbf{verified by tests of significance}) are highlighted in bold. 
    Best-performing models are colored in gray background.
    }
    \centering
    \vspace{-5pt}
    \begin{adjustbox}{width=0.99\textwidth}
    \begin{tabular}{c cccccccccccccccccc}
    \toprule
    & \multicolumn{2}{c}{ASVspoof (All)} &  \multicolumn{15}{c}{CodecFake} \\  
    \cmidrule(lr){2-3}
    \cmidrule(lr){4-18}
    & AASIST & AASIST-L & A & B1 & B2 & B3 & C & D1 & D2 & D3 & E & F1 & F2 & F3 & F4 & F5 & F6 \\ 
    \midrule
    
    $\text{VALL-E}^*$   & 9.93 & 13.91 & \textbf{0.53} & 31.79 & 35.50 & 58.94 & \textbf{5.30} & \textbf{4.50} & 16.95 & \textbf{2.65} & \cellcolor{gray!20}{{\textbf{0.40}}} & \textbf{1.99} & 32.72 & 41.85 & 21.06 & 42.52 & 56.29 \\
    $\text{VALL-E X}^\dagger$ & 36.11 & 33.33 & 25.00 & 52.78 & 58.33 & 50.00 & 41.67 & \textbf{11.11} & 22.22 & \textbf{13.89} & \cellcolor{gray!20}{\textbf{5.56}} & 22.22 & 44.44 & 41.67 & 50.00 & 50.00 & 58.33 \\
    $\text{SpeechX}^\dagger$  & 33.33 & 25.00 & 50.00 & 66.67 & 58.33 & 58.33 & 41.67 & 16.67 & 50.00 & 41.67 & \cellcolor{gray!20}{8.33} & 16.67 & 41.67 & 50.00 & 50.00 & 50.00 & 50.00 \\
    \bottomrule
    \multicolumn{18}{l}{ \footnotesize * Implementation from https://huggingface.co/amphion/valle\_librilight\_6k; $\dagger$ Audio collected from respective demo pages.}
    \vspace{-6mm}
    \end{tabular}
    \end{adjustbox}
    \label{tab: tts}
\end{table*}

\section{Experimental setup}
\label{setup}

\subsection{Implementation details}
We adopt AASIST-L~\cite{aasist} \footnote{https://github.com/clovaai/aasist}, a state-of-the-art deepfake detection model, as our backbone model to train the anti-spoofing models.
We choose AASIST-L rather than AASIST as the backbone, because AASIST-L is less prone to overfitting.
This model architecture is leveraged to get well-trained models on both the ASVspoof and CodecFake data.
For the ASVspoof data, we utilize the author's released checkpoints, as they have already conducted a thorough hyperparameter search.
For CodecFake data, we do a careful hyperparameter search for the batch size and learning rate, to achieve optimal results.
The final learning rate is 0.001 with an Adam optimizer, and the batch size is 16.
During training, we combine one of the 15 CodecFake subsets with the corresponding original VCTK utterances as training data.
Then, we employ the respective CodecFake validation subset for model selection.

\subsection{Evaluation details}
We consider four evaluation sets:
(1) ASVspoof 2019 evaluation set;
(2) CodecFake evaluation set;
(3) VCTK evaluation set synthesized using an open-source reimplementation of VALL-E \cite{wang2023neural} \footnote{https://huggingface.co/amphion/valle\_librilight\_6k};
(4) Synthesized and ground truth audio samples collected from the demo pages of codec-based speech synthesis systems, including VALL-E X \cite{zhang2023speak} \footnote{https://www.microsoft.com/en-us/research/project/vall-e-x/} and SpeechX \cite{wang2023speechx} \footnote{https://www.microsoft.com/en-us/research/project/speechx/}.
We resort to collecting audio samples from VALL-E X and SpeechX demo pages 
because they do not release pretrained models for inference due to concerns about potential malicious use. 
Potential malicious deepfake attacks may even impact open-source efforts, underscoring the significance of our research. 
Once deepfake attacks become detectable, researchers may become more willing to release their models.
For demo page audio samples, we concentrate solely on tasks that generate speech with minimal noise, excluding audio generation and samples with significant distortion. 
Our evaluation primarily focuses on samples intended for speaker-conditioned text-to-speech synthesis, speaker-conditioned speech-to-speech translation, speech editing, voice conversion, etc.
We collect pairs of real and fake speech.
Real speech comes from the ground truth or the speaker prompt,
whereas model-generated speech is categorized as fake speech.
The statistics of the four evaluation sets used for testing are shown in the \textit{Zero-shot Evaluation} part of Table~\ref{tab: testing_data_info}.
We adopt equal error rate for evaluation, as it is the mainstream metric to evaluate anti-spoofing models \cite{todisco2019asvspoof}.

\section{Experimental results}
\label{result}

\subsection{Detecting speech generated by neural audio codecs}
\label{subsec: codec}
The cross-testing results (EERs. Lower is better) are shown in Figure~\ref{fig: codecs_t}.
The rows denote the training data subsets (CodecFake and ASVspoof), and the columns denote the CodecFake testing data subsets.
We have the following observations:
\begin{itemize}
    \item Models (AASIST and AASIST-L) trained on the ASVspoof dataset exhibit significantly high EERs (usually higher than 30\%) across most of the CodecFake testing subsets. 
    This suggests that models trained on ASVspoof struggle to detect synthesized speech from codec models. 
    \item Typically, training on a specific CodecFake training subset leads to the lowest EER for the corresponding CodecFake testing subset (e.g., row A as training and column A as testing; row B1 as training and column B1 as testing). 
    Training on a particular type of CodecFake training subset generally results in relatively low EERs for the corresponding type of CodecFake testing subset (e.g., row B1 as training, column B1, B2 and B3 as testing). 
    A possible reason is that the in-distribution testing conditions will perform better than out-of-distribution testing conditions.
    \item The model trained on training subset A (row A) demonstrates good generalizability to most of the testing subsets (columns A to F6). In SpeechTokenizer (A) \cite{zhang2023speechtokenizer}, authors introduce a technique to allocate different types of information across different layers of codebooks. This disentanglement of information may contribute to the above generalizability.
\end{itemize}

\subsection{Detecting speech generated by codec-based speech synthesis systems}
\label{subsec: tts}
Table~\ref{tab: tts} illustrates the testing results for models trained by different kinds of training datasets.
The columns denote which training set is adopted for training. 
For ASVspoof, all the training data are adopted to train two model variants, AASIST and AASIST-L.
For the CodecFake dataset, each column denotes which training subset is adopted to train the AASIST-L model.
The rows denote the sources of testing sets.
We have the following observations:
\begin{itemize}
    \item 
    Models trained with the ASVspoof dataset do not yield satisfactory performance on the three codec-based speech synthesis systems. 
    This is likely because ASVspoof is comprised of speech generated by traditional speech synthesis models, which differ significantly from current codec-based systems. 
    This underscores the urgent need for a codec-based deepfake detection dataset.
    \item 
    VALL-E claims to be able to replicate anyone's voice with just three seconds of speech, which raises serious concerns about potential malicious impersonation.
    However, we find that model trained from CodecFake train subsets A and E effectively detect fake speech generated by VALL-E, with remarkably low equal error rates (0.53\% and 0.4\%). 
    For demo samples from VALL-E X and SpeechX, models trained on CodecFake train subset E achieve the best EER. 
    We also note that models trained on subsets D1 and F1 also exhibit better detection performance than ASVspoof baselines.
    \item 
    We perform statistical tests to verify results in Table~\ref{tab: tts} are statistically significant.
    For each testing source, we compare classifier predictions from models trained on each CodecFake train subset to prediction from the best performing ASVspoof classifier.
    With a null hypothesis of identical equal error rates, we conduct two tests, paired sample t-test and Wilcoxon signed-rank test, using a p-value threshold of 0.01.
    For VALL-E and VALL-E X, models highlighted in bold significantly outperform the classifiers trained on ASVspoof.
    For SpeechX, due to the small number of data samples, we cannot verify whether gains achieved are meaningful.
\end{itemize}

\vspace{-3mm}
\section{Conclusion}
\label{Conclusion}
In this study, we assembled a comprehensive collection of contemporary codec models, resulting in the creation of the CodecFake dataset. 
Our extensive experiments verify that while anti-spoofing models trained on the ASVspoof dataset struggle to detect synthesized speech from codec models and current codec-based speech generation models, our CodecFake dataset significantly enhances anti-spoofing models' capability to do so. 
We achieved an equal error rate of 0.4\% for synthesized speech from VALL-E, a leading codec-based speech generation model.
We will release all codes and data, aiming to contribute to mitigating malicious attacks from current advanced codec-based speech synthesis systems.

\noindent\textbf{Acknowledgement}:
We thank the National Center for High-performance Computing (NCHC) of National Applied Research Laboratories (NARLabs) in Taiwan for providing computational and storage resources.

\newpage
\bibliographystyle{IEEEtran}
\bibliography{mybib}

\end{document}